\documentclass[twocolumn,aps,prl,superscriptaddress]{revtex4-2}
\usepackage[utf8]{inputenc}
\usepackage{amsmath}
\usepackage{stix2}
\usepackage{bm}
\usepackage{xcolor}
\usepackage{graphicx}

\begin{document}

 \title{Kohn--Luttinger Superconductivity in Flat Chern Bands}

\author{Nianlong \surname{Zou}}
\affiliation{Department of Physics and Astronomy, University of Tennessee, Knoxville, TN 37996, USA}

\author{Cheng Xu}
\affiliation{Max Planck Institute for Chemical Physics of Solids, 01187, Dresden, Germany}

\author{Yafis Barlas}
\affiliation{Department of Physics, University of Nevada, Reno, Nevada 89557, USA}

\author{Yang Zhang}
\affiliation{Department of Physics and Astronomy, University of Tennessee, Knoxville, TN 37996, USA}
\affiliation{Min H. Kao Department of Electrical Engineering and Computer Science, University of Tennessee, Knoxville, Tennessee 37996, USA}

\begin{abstract}
Recent observations of superconductivity near correlated topological phases in flat bands suggest a facile link between flat-band geometry and electron pairing. In this work, we reveal a geometry-driven Kohn--Luttinger mechanism in which Landau-level-like form factors align the attractive lobe of the RPA-screened Coulomb interaction with the form-factor peak, generating an anomalously strong attractive channel near local band extrema. Using the Skyrmion lattice model as a minimal realization, we show that for spin-unpolarized pairing the form-factor magnitude enforces an emergent momentum-space translational symmetry and selects an extended-$s$ instability concentrated at small Fermi pockets, while for spin-polarized pairing the form-factor phase drives chiral $p$- and $f$-wave order without invoking spin fluctuations. The band-extrema enhancement persists in higher Landau-level analogs and survives finite-temperature screening and Berezinskii--Kosterlitz--Thouless phase fluctuations. Our work establishes quantum geometry as a key organizing principle for unconventional pairing in flat Chern bands.
\end{abstract}

\maketitle

\textbf{Introduction.---}Flat Chern bands have become a central platform for correlated topological phases of matter. The experimental realization of fractional Chern insulators (FCIs) in moir\'e superlattices has renewed interest in the interplay between correlated electrons and nontrivial topology~\cite{Cai2023, Park2023, Zeng2023, Xu2023,Morales-Duran2024}. A key ingredient for these exotic phases is the presence of a flat Chern band that mimics the quantum geometry of the Landau level~\cite{Wang2021, Ledwith2020, Parameswaran2013, Roy2014, Claassen2015}. Beyond the FCI regime, recent experiments have unveiled robust superconducting domes in the immediate vicinity of these phases~\cite{Xu2025, Wang2025, Xu2025a, Zhou2021, Dong2024, Jahin2026}. The natural question is whether the quantum geometry that stabilizes the FCIs also reshapes the pairing instability.

Given that the adjacent FCI phase is driven exclusively by Coulomb repulsion, the proximate superconductivity is likely of electronic origin as well. One natural candidate is the Kohn--Luttinger (KL) mechanism~\cite{Kohn1959,Luttinger1966, Kagan1988, Chubukov1993,Gonzalez2019}, where purely repulsive interactions generate an effective attraction via screening. The conventional KL mechanism relies on the $2k_F$ Lindhard cusp to generate Friedel oscillations~\cite{Shankar1994, Raghu2010}, typically yielding high-angular-momentum pairing confined to the Fermi surface. Neither feature carries over unchanged to a flat Chern band. First, the flat-band nature dictates that pairing is no longer confined to the Fermi surface, but instead involves the collective participation of Bloch states across the entire Brillouin zone. For example, instead of the conventional Fermi-surface contribution, the superfluid stiffness is determined by the quantum metric integrated over the whole Brillouin zone~\cite{Peotta2015, Liang2017, Hu2019, Xie2020,Jiang2024}. Second, nontrivial geometric factors reshape the effective interaction itself, rendering chiral superconductivity and finite-momentum pairing accessible within intravalley pairing channels~\cite{Guerci2025,Shavit2025, Yu2025, Julku2020, Peri2021, Chen2026}. So far, this enhancement has been analyzed within a single parabolic band through Fermi-surface quantities at fixed density~\cite{Jahin2026, Shavit2025, Guerci2025}; its fate in a genuine lattice band, and its filling dependence, remain open.

In this work, we reveal a band-extrema enhancement of Kohn--Luttinger pairing in flat Chern bands, using the Skyrmion lattice model (SLM)~\cite{Wu2019,Ledwith2020,Khalaf2021,Paul2023} as a minimal paradigm for ideal flat Chern bands. Depending on the presence or absence of time-reversal symmetry, we identify two distinct pairing paradigms. In the spin-unpolarized regime, an emergent momentum-space translational symmetry drives the RPA-screened interaction towards a unique plane-wave pairing channel, rather than conventional high-angular-momentum channels. In contrast to previous flat-band theories that assume a uniform attraction and predict optimal pairing near half-filling~\cite{Peotta2015, Julku2016, Hofmann2023}, we find that the RPA-generated attraction is strongly filling-dependent and peaks at small Fermi pockets, anomalously enhancing the superconducting $T_c$ near local band extrema. The mechanism extends to spin-polarized systems with chiral pairing and to higher-Landau-level-like bands, and persists under thermal smearing of RPA screening and Berezinskii--Kosterlitz--Thouless (BKT) phase fluctuations.

\textbf{Skyrmion lattice model.---}The SLM describes itinerant electrons coupled to a periodic Skyrmion texture~\cite{Nagaosa2013, Hamamoto2015}:
\begin{equation}
H^\sigma = \frac{\bm{p}^2}{2 m} + \sigma \,\hat{\boldsymbol{s}} \cdot \bm{J}(\bm{r}),
\end{equation}
where $m$ denotes the effective mass, $\hat{\boldsymbol{s}}$ represents the Pauli matrices acting on the pseudospin degree of freedom, and $\bm{J}$ is the skyrmion texture. $\sigma=\pm1$ represents different spin channels, which correspond to the valley degree of freedom in moir\'e systems due to spin-valley locking. In the adiabatic limit, the noncoplanar Skyrmion texture induces a Berry connection that acts as a magnetic field, mapping the system onto a Landau-level problem~\cite{Hamamoto2015}. As illustrated in Fig.~1(a), the model realizes a set of flat Chern bands with Chern number $C=\sigma$ for each spin channel. These bands exhibit an extremely narrow bandwidth, rendering the effective interaction the dominant energy scale in the system. Fig.~1(b) further analyzes the magnitude of form factor $|\Lambda_{\bm{k}, \bm{k}+\bm{q}}|$. The numerical results (blue dots) show excellent agreement with the analytical form factor of the lowest Landau level (LLL): $|\Lambda^{\mathrm{LLL}}_{\bm{q}}| = e^{-l_B^2 |\bm{q}|^2 / 4}$, where $l_B$ denotes the magnetic length. Crucially, its negligible dependence on momentum $\bm{k}$ underpins the emergent momentum-space translational symmetry and motivates the analytical approximations below.

\begin{figure}
    \includegraphics[width=\linewidth]{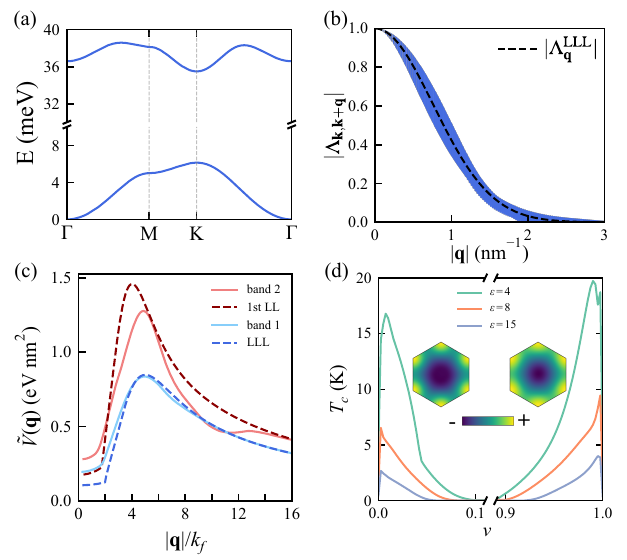}
    \caption{
    (a) Band structure of the SLM. (b) Comparison of form factors between the SLM and ideal Landau levels. The blue markers indicate the magnitude of the SLM form factor, while the black dashed line represents the corresponding ideal LLL form factor. (c) The RPA-screened potential $\tilde{V}(\mathbf{q})$ of the SLM compared with Landau levels. Results are shown for the first band ($\nu=0.05$, corresponding to the LLL) and the second band ($\nu=1.05$, corresponding to the first LL), with a relative permittivity of $\epsilon_r=10$. (d) Dependence of the  superconducting transition temperature $T_c$ on the filling factor $\nu$ for various relative permittivities $\epsilon_r$. The plot covers filling ranges near the bottom ($\nu \approx 0$) and top ($\nu \approx 1$) of the band.
    }
    \label{fig1}
\end{figure}

Given the well-isolated nature of moir\'e minibands [Fig.~1(a)], KL superconductivity is governed by the effective interaction projected onto this specific band:
\begin{equation}
V^{\bm{\sigma,\sigma^\prime}}_{\bm{k,k'}}=\sum_{\bm{GG^\prime}}	\tilde{V}_{\bm{G,G^\prime}}(\bm{k-k'}) \Lambda^{\sigma}_{\bm{k,k'-G}}\Lambda^{\sigma^\prime}_{\bm{-k,-k'+G^\prime}}.\label{vkkp}
\end{equation}
Here, the effective interaction $\tilde{V}$ is evaluated within the random phase approximation~\cite{Gell-Mann1957, Bohm1957,Adler1962} by summing infinite-order bubble diagrams: $[\tilde{V} (\bm{q})]^{- 1}_{\bm{G, G'}} = V^{-1}_{0} (\bm{q+G})\delta_{\bm{G, G'}} - \Pi_{\bm{G, G'}} (\bm{q})$. The bare potential is modeled as a dual-gate screened Coulomb potential, $V_0(\bm{q})= e^2\tanh(|\bm{q}|d_s)/(2\epsilon_0\epsilon_r|\bm{q}|)$, with a gate distance $d_s = 10\,\mathrm{nm}$. Near the band extrema, by approximating the dispersion as quadratic, $\epsilon_{\bm{k}} \approx \bm{k}^2/(2m^*)$, and approximating the exact form factor with $\Lambda^{LLL}$, the Lindhard function takes the following analytical form:
\begin{equation}
    \Pi_{\bm{G},\bm{G'}}(\bm{q}) = g\delta_{\bm{G},\bm{G'}}|\Lambda^{\mathrm{LLL}}_{\bm{q}+\bm{G}}|^2\Pi_{\mathrm{2D}}(\bm{q}+\bm{G}).\label{PiGG}
\end{equation}
Here $g$ denotes the spin degeneracy factor, with $g=2$ ($1$) for spin-unpolarized (spin-polarized) cases. This expression describes the 2D free-electron Lindhard function $\Pi_{\mathrm{2D}}$ modulated by the LLL form factor. As shown in Fig.~2(a), this approximation agrees well with numerical SLM calculations.

Due to RPA overscreening, an effective attraction is generated in $V^{\bm{\sigma,\sigma^\prime}}_{\bm{k,k'}}$, leading to a superconducting instability. Within the generalized BCS framework, the critical temperature $T_c$ and order parameter $\Delta^{\sigma, \sigma'}_{\bm{k}}$ are obtained by solving the linearized gap equation:
\begin{equation}
\Delta^{\sigma, \sigma'}_{\bm{k}} = - \frac{1}{A}\sum_{\bm{k'}} V^{\sigma, \sigma'}_{\bm{k, k'}} I^{\sigma,\sigma'}_{\bm{k'}} \Delta^{\sigma, \sigma'}_{\bm{k'}},\label{gapeq}
\end{equation}
with
\begin{equation}
I^{\sigma, \sigma'}_{\bm{k}} = \frac{1}{2(\epsilon_{\bm{k}}^{\sigma} + \epsilon_{\bm{- k}}^{\sigma'})}\left[ \tanh \frac{\beta \epsilon_{\bm{k}}^{\sigma}}{2} + \tanh\frac{\beta \epsilon_{\bm{- k}}^{\sigma'}}{2} \right].\label{Ikernel}
\end{equation}
We first focus on the spin-unpolarized case, where time-reversal symmetry facilitates pairing between opposite flavors ($\sigma^\prime = -\sigma$). Consequently, the effective interaction in Eq.~\eqref{vkkp} depends solely on the form factor magnitudes $|\Lambda^{\sigma}_{\bm{k,k'-G}}|$. The numerical results reveal that the SLM departs significantly from the conventional BCS paradigm: $T_c$ is anomalously enhanced at the band edges across a wide range of dielectric constants $\epsilon_r$ [Fig.~1(d)]. The resulting order parameter remains nearly invariant between high- and low-filling regimes despite their distinct Fermi surfaces, highlighting that the pairing is a collective Brillouin-zone effect rather than a Fermi-surface response.

\textbf{RPA interaction in the plane-wave channel.---}To elucidate the mechanism, we simplify the problem by exploiting emergent symmetries and the flat-band limit, which allows for an analytical solution to Eq.~\eqref{gapeq}. First, the flat-band limit allows us to neglect the $\bm{k}$ dependence of $I(\bm{k})$. Within this approximation, the order parameter corresponds directly to the lowest eigenstate of $V^{\bm{\sigma,-\sigma}}_{\bm{k,k'}}$, representing the channel with the strongest effective attraction. An emergent $\bm{k}$-space translational symmetry arises in $V^{\bm{\sigma,-\sigma}}_{\bm{k,k'}}$ from two features near the band extrema: (i) the off-diagonal contributions of $\tilde{V}_{\bm{G},\bm{G}^\prime}$ are negligible due to Eq.~\eqref{PiGG}; and (ii) the magnitudes of the form factors are independent of $\bm{k}$. As a result, $V^{\bm{\sigma,-\sigma}}_{\bm{k,k'}}$ depends solely on the momentum transfer $\bm{q}=\bm{k}-\bm{k}^\prime$:
\begin{equation}
V^{\bm{\sigma,-\sigma}}_{\bm{k,k'}} \approx \sum_{\bm{G}}\tilde{V}_{\bm{G,G}}(\bm{q}) e^{-l_B^2 |\bm{q}+\bm{G}|^2 / 2}.\label{vkkp_approx}
\end{equation}
The validity of approximating $V^{\bm{\sigma,-\sigma}}_{\bm{k,k'}}$ by Eq.~\eqref{vkkp_approx}, combined with the flat-band approximation described above, is numerically verified in Fig.~2(a). Despite a slight reduction in $T_c$, the characteristic band-extrema enhancement is preserved, and the pairing symmetry remains unchanged.

\begin{figure}[t]
    \includegraphics[width=\linewidth]{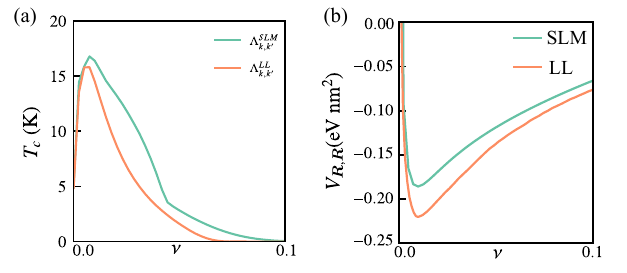}
    \caption{
    (a) $T_c$ curves obtained with the exact SLM form factor and the Landau level approximation, both using the SLM band structure to preserve lattice periodicity. (b) Effective attraction strength in the plane-wave channel versus filling. Both the SLM numerical results and the analytical solution obtained from Eq.~\eqref{PiGG} exhibit a pronounced peak near the band edge. Note that the apparent discrepancy, where the SLM yields a higher $T_c$ in (a) despite a weaker attraction in (b), arises because the analytical LL curve in (b) assumes free-electron bands, whereas both calculations in (a) strictly use the SLM dispersion.
    }
    \label{fig2}
\end{figure}

Given the $\bm{k}$-space translational invariance, $V_{\bm{k}, \bm{k}'}$ becomes diagonal in the conjugate real-space basis defined by the lattice vectors $\bm{R}$: $V_{\bm{R},\bm{R}'}=(1/N_k) \delta_{\bm{R},\bm{R}'}  \sum_{\bm{q}} V_{0,\bm{q}} e^{-i\bm{R}\cdot\bm{q}}$. We find that the strongest attractive channels correspond to the six nearest-neighbor vectors $\bm{R}_0$ related by $C_6$ symmetry. These channels possess equal attractive strengths, forming a sixfold-degenerate interaction sector. Fig.~\ref{fig2}(b) illustrates the striking filling dependence of this effective attraction. Both the SLM numerical results and the analytical predictions exhibit a pronounced maximum near the band edge. We attribute this enhancement to a geometric alignment effect: near the band extrema, the small Fermi momentum ensures small momentum transfer, which aligns the attractive dip of the RPA potential directly with the maximum of the geometric form factor. This alignment maximizes the effective attractive potential available for pairing, confirming that the band-edge $T_c$ enhancement arises from the strong attraction in this plane-wave channel. This geometric enhancement holds across a wide range of parameters and models, including the Haldane model (see Supplemental Material for details~\cite{SM}).


Turning to the order parameter, numerical solutions of the gap equation reveal an even-parity leading instability, dictating a spin-singlet pairing state. To analytically derive this order parameter, we project the effective interaction onto the threefold-degenerate subspace spanned by the even-parity basis functions $\cos(\bm{k}\cdot\bm{R}_i)$. Accounting for weak translational-symmetry breaking introduces off-diagonal inter-bond scattering, reducing the interaction to an effective $3 \times 3$ matrix:

$$V_{\bm{R},\bm{R}'} = \begin{pmatrix}
V_0 & \delta_V & \delta_V \\
\delta_V & V_0 & \delta_V \\
\delta_V & \delta_V & V_0
\end{pmatrix}$$
This interaction matrix admits two fundamental solutions: an extended $s$-wave state with eigenvalue $V_0 + 2\delta_V$ and a doubly degenerate $d$-wave state with eigenvalue $V_0 - \delta_V$. Given the strong attractive interaction, the off-diagonal terms are typically negative; consequently, the $C_3$-invariant solution is energetically favored, yielding the order parameter depicted in Fig.~1(d). While both singlet and triplet pairing channels are possible for the spin unpolarized state, the even-parity orbital structure favors a spin-singlet order parameter.

\begin{figure}
    \includegraphics[width=\linewidth]{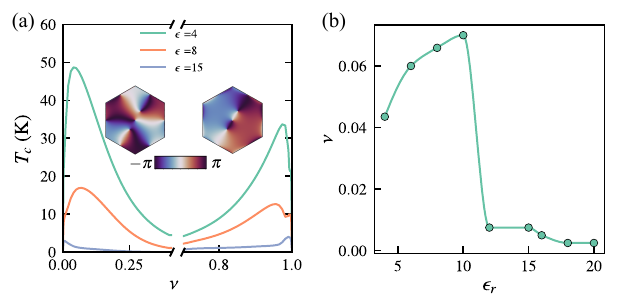}
    \caption{
    (a) Dependence of $T_c$ and the order parameter phase on filling for the spin-polarized skyrmion model. The pairing symmetry evolves from chiral $f$-wave in the low-filling region to chiral $p$-wave at high filling. (b) Variation of the optimal filling with respect to the dielectric constant $\epsilon_r$.
    }
    \label{fig3}
\end{figure}

\textbf{Spin-polarized chiral pairing and higher bands.---}We further investigate the spin-polarized case, where the system's time-reversal symmetry is spontaneously broken. As illustrated in Fig.~3(a), our calculations reveal a superconducting phase with even higher $T_c$ and a broader filling range than the spin-unpolarized case. The anomalous $T_c$ enhancement near the band edge persists in this channel. However, the order parameter in this case is distinct due to the absence of time-reversal symmetry. In contrast to the spin-unpolarized case, the effective interaction $V^{\sigma,\sigma}_{\bm{k},\bm{k}'}$ is now sensitive not only to the magnitude of the form factor but also to its momentum-dependent phase. Consequently, the emergent translational symmetry is broken and replaced by a pseudomagnetic translational symmetry in $\bm{k}$-space~\cite{SM}. As a result, chiral superconductivity is generated, featuring a chiral $f$-wave state in the low-filling region and a chiral $p$-wave state in the high-filling region. The optimal filling exhibits a non-monotonic dependence on the dielectric constant $\epsilon_r$ [Fig.~3(b)].



We also extended our analysis to the second band~\cite{SM}, where both the band structure and form factors differ significantly from the lowest band and the mechanism can be generalized. As shown in Fig.~1(a), the second band features a local extremum at the $\Gamma$ point, which leads to a $T_c$ enhancement near $\nu=1.11$. This finding extends our conclusion from a specific ``band-edge enhancement'' to a broader ``band-extrema enhancement.'' Regarding the order parameter, distinct behaviors emerge: the strongest attraction channel shifts from the nearest-neighbor to the next-nearest-neighbor, although the extended $s$-wave solution remains favored. As detailed in the Supplemental Material~\cite{SM}, this transition originates from the interplay between 1LL form factors favoring longer-range interactions and the order parameter's structural realignment to the flattest band regions at the additional extrema.

\begin{figure}
    \includegraphics[width=\linewidth]{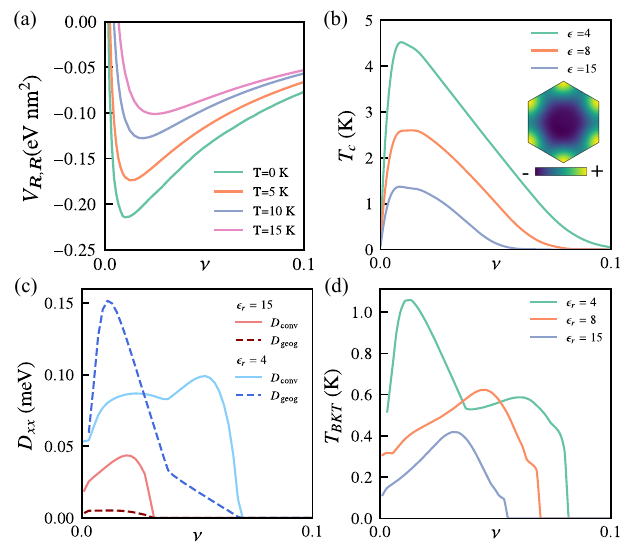}
    \caption{
    (a) RPA interaction strength in the plane-wave channel as a function of filling at different temperatures. (b) $T_c$ versus filling, accounting for the effect of temperature on RPA screening. Inset: Order parameter, which remains nearly invariant with filling. (c) Conventional and geometric contributions to the superfluid weight versus filling for various $\epsilon_r$. (d) BKT transition temperature. For small $\epsilon_r$, a band-edge enhancement is observed, driven by the dominant geometric contribution. 
    }
    \label{fig4}
\end{figure}


\textbf{Thermal and phase fluctuations.---}The enhanced $T_c$ near the band edge is comparable to the bandwidth of the SLM. This indicates that the temperature dependence of the RPA screening must be explicitly considered, particularly in systems with a small dielectric constant where Coulomb interactions are prominent. At zero temperature, the Lindhard cusp causes a rapid suppression of screening for $q > 2k_F$, creating a strong momentum dependence in the effective potential $V(q)$ that drives the pairing attraction. At finite temperatures, however, thermal smearing rounds off this $2k_F$ singularity, smoothing out the momentum dependence of the screening response. Consequently, as temperature increases, the attraction weakens and the attractive peak shifts away from the band edge, as illustrated in Fig.~4(a). Although $T_c$ is reduced compared to the zero-temperature approximation, the band-edge enhancement persists. Furthermore, the pairing symmetry is preserved, with the order parameter consistently preferring the $C_3$-invariant channel.

To capture the macroscopic phase coherence in our two-dimensional system, we evaluate the BKT transition temperature, governed by the universal relation $k_B T_{\mathrm{BKT}} = (\pi/8)  D(T_{\mathrm{BKT}})$. In a single-band system, the superfluid stiffness $D(T)$ comprises two parts: a conventional, dispersion-driven term ($D_{\mathrm{conv}}$) and a geometric contribution ($D_{\mathrm{geo}}$) derived from the intraband quantum metric. Evaluating $D(T)$ below the mean-field transition requires solving the fully nonlinear gap equation, which amounts to substituting the bare band dispersion in $I(\bm{k})$ of Eq.~\eqref{Ikernel} with the quasiparticle excitation energy $E_{\bm{k}} = \sqrt{\xi_{\bm{k}}^2 + |\Delta_{\bm{k}}|^2}$. Typically, the nonlinearity of $E_{\bm{k}}$ generates higher-order spatial harmonics in the order parameter. However, the interaction $V_{\bm{R},\bm{R}'}$ is nearly diagonal, and the effective attraction is only pronounced at the nearest-neighbor shell $\bm{R}_0$, while the primary repulsive channel remains on-site ($V_{00}$). Consequently, the higher-order harmonics are suppressed, and the finite-temperature order parameter is restricted to the form $\Delta(T) = \alpha(T)\Delta(T_c) + \beta(T)\Delta_0$~\cite{SM}. This ensures that the emergent plane-wave pairing symmetry identified at $T_c$ is preserved down to the BKT transition.

The numerical results for the superfluid stiffness and $T_{\mathrm{BKT}}$ are presented in Fig.~4(c,d). The origin of the superfluid weight undergoes a crossover depending on the dielectric environment. In the strong-screening (high-$\epsilon_r$) regime, $D_{\mathrm{conv}}$ dictates the superconducting response, exhibiting a non-monotonic filling dependence that initially increases away from the band edge. Conversely, in the weakly screened (low-$\epsilon_r$) regime, the topological flat-band nature takes precedence, and the response is governed by $D_{\mathrm{geo}}$, which decays monotonically with filling. Computing $T_{\mathrm{BKT}}$ from the total stiffness confirms that the $T_c$ enhancement near the band extrema survives phase fluctuations, yielding a macroscopically coherent superconducting phase.

We note that spin- or valley-polarized Chern insulating states, naturally favored by repulsive interactions in flat bands, may further promote superconductivity through a spin-fluctuation mechanism~\cite{RevModPhys.84.1383}. Such fluctuations would favor the same spin-polarized channel that already displays the larger $T_c$ in our geometry-driven KL mechanism, suggesting a broader interplay between magnetic order and quantum geometry that we leave for future work.

\textbf{Conclusion.---}Our analysis identifies a geometric alignment principle: in flat Chern bands with Landau-level form factors, the small Fermi pockets near band extrema bring the strong screening at small $k_F$ into alignment with the form-factor peak, sharply enhancing RPA-mediated pairing. This principle organizes the spin-unpolarized and spin-polarized channels into a single mechanism set by form-factor magnitude and phase, respectively. A direct experimental consequence is that the optimal $T_c$ in any ideal flat Chern band lies near the local band extrema, providing a design rule for moir\'e superconductors proximate to fractional Chern insulators.

\begin{acknowledgments}

\end{acknowledgments}

\bibliography{reference}

\end{document}